# Virucidal Efficacy of Laser-Generated Copper Nanoparticle Coatings Against Model Coronavirus and Herpesvirus

Shahd Bakhet[1], Rasa Mardosaitė[1], Mohamed Ahmed Baba[1], Asta Tamulevičienė[1,2], Brigita Abakevičienė[1,2], Tomas Klinavičius[1], Kristupas Dagilis[2], Simas Račkauskas[1], Sigitas Tamulevičius[1,2], Raimundas Lelešius[3,4], Dainius Zienius[3,4], Algirdas Šalomskas[4], Tomas Tamulevičius[1,2]*

[1]Institute of Materials Science of Kaunas University of Technology, K. Baršausko St. 59, LT-51423, Kaunas, Lithuania

[2]Department of Physics, Kaunas University of Technology, Studentų St. 50, LT-51368, Kaunas, Lithuania

[3]Department of Veterinary Pathobiology, Lithuanian University of Health Sciences, Tilžės St. 18, LT-47181 Kaunas, Lithuania

[4]Institute of Microbiology and Virology, Lithuanian University of Health Sciences, Tilžės St. 18, LT-47181 Kaunas, Lithuania

*Corresponding author: T. Tamulevičius, tomas.tamulevicius@ktu.lt, Tel: +370 (37) 313432, Institute of Materials Science of Kaunas University of Technology, K. Baršausko St. 59, LT-51423, Kaunas, Lithuania

## Abstract

High-efficiency antiviral surfaces can be effective means to fight against viral diseases such as the recent Covid-19 pandemic. Copper and copper oxides, as well as their nanoparticles (Cu NPs) and coatings, are among the effective antiviral materials having internal and external biocidal effects on viruses. In this work, stable Cu nanoparticle colloids were produced via femtosecond laser ablation of the metal target in water containing sodium citrate. Raman spectroscopy and X-ray diffraction studies confirmed that the 32 nm mean size nanoparticles are mixtures of mainly metallic copper and copper (I) oxide $Cu_2O$. Polyvinyl butyral was utilized as the binding agent for the spray-coated Cu NPs. The virucidal efficacy of such coatings containing different Cu content ranging from 2.9 at.% to 11.2 at.%. was confirmed against animal-origin coronavirus containing RNA, the agent of avian infectious bronchitis (IBV), and herpesvirus containing DNA, the agent of bovine herpesvirus (BoHV-1) infection. It was demonstrated that after a short time of exposure, the Cu NPs-based coatings do not have a toxic effect on the cell cultures while demonstrating a negative effect on the biological activity of both model viruses that was confirmed by quantification of the viruses via the determination of tissue culture infectious dose ($TCID_{50}$) virus titre and their viral nucleic acids via determination of threshold cycle (Ct) employing real-time polymerase chain reaction analysis. The assays showed that the decrease in $TCID_{50}$ virus titre and increase in Ct values correlated with Cu content in Cu NPs-based coatings for both investigated viruses. Contact with coatings decreased IBV and BoHV-1 numbers from 99.42% to 100.00% and from 98.65% to 99.96% respectively. These findings suggest that Cu NPs show inhibitory effects leading to the inactivation of viruses and their nuclei regardless of the presence of a viral envelope.

## 1. Introduction

According to the World Health Organization (WHO) close to 700 million people had Covid-19 and close to 7 million related deaths were counted at the beginning of the year 2023 [1]. Having in mind the overall population exceeding 8 billion [2] one can estimate that over 8% of people suffered from the disease and statistically more than one vaccine was applied per person as over 13 billion vaccine doses were administered. Although Covid-19 is no longer a pandemic disease, more than 20 000 cases are still registered every day around the world. One of the ways to deal with the outbreaks is good hygiene, which can be implemented by the greater use of antiviral surfaces effective against ribonucleic acid (RNA) like viruses such as coronavirus and influenza virus [3]. Specialized assays have emerged for the evaluation of viral resistance [4, 5]. The studies showed, that there are significant differences in the biophysical stability factors of these viruses on the surfaces of different materials (plastic, stainless steel, cardboard, copper), and this led the researchers to investigate different hypotheses to evaluate the physical and chemical properties of the surfaces of various coatings (including on face masks) for possibly reducing the stability of SARS-CoV-2 and thus its potential transmission to healthy people. Several key properties that affect the survival of viruses on surfaces were identified [6, 7]. The first one is the shape and type of surface of the contact material, such as porosity and roughness. The second is physical abiotic factors that affect the surface of the material, such as relative humidity, temperature, and exposure to light (rays of different solar spectra). The third is chemical, such as pH, the presence of reactive ions, the state of adsorption, and the presence of organic matter or specific chemical components in the environment. It should be emphasized that many viruses, depending on their external biochemical structure, can be stabilized and protected by surrounding organic matter, such as saliva or mucus droplets. In this case, the presence of substances and external structures such as bacteria biofilms, fats, proteins, or carbohydrates in the contact environment of the virus can additionally and significantly increase their structural stability and resistance [6, 7].

Different combinations of metal oxide materials concerning biological activity were studied in different biosystems, but silver and copper derivatives and their compounds were among the effective ones even though the means of their internal and external virucidal activity mechanisms against viruses are still not fully understood. At the same time, metal ions are an integral part of some viral proteins and play an important role in their survival and pathogenesis [8].

Cu is an indispensable trace element for a wide range of biological functions in all living systems and due to its potent antiviral and antioxidant properties, it may be promising even for prophylactic modalities against COVID-19 infection [9]. According to the National Institute of Health (NIH), the current recommended Cu dietary allowance (RDA) for adult females and males is 900 µg/day [10]. Copper supplementation is recommended to accompany zinc supplementation [11].

In recent years there has been an increasing interest in the use of self-disinfecting surfaces to reduce the concentration of different pathogens for preventing the spread of diseases. Various materials were used as additional safety measures, for example, surface coatings with ZnO [12] and CuO nanoparticles (NPs) [13]. Due to its effective contact denaturing effect and price [14], Cu is the most used metal for the development of antimicrobial surfaces. It can be used pure, alloyed, in composites with different structures, and NPs can be used in a variety of products such as door handles, railings, and textiles [15, 16]. Various forms of Cu compounds, especially copper oxide nanoparticles (CuO-NPs), are one of the most popular in the field of antimicrobial research. These NPs have been widely adopted in industry [17], due to their versatile physical properties and low cost of production [18]. In the past, several research projects have emphasized the inhibitory effect of CuO-NPs on viruses [19], but most studies have focused on the effectiveness of CuO against bacteria [18, 20].

Cu NPs can be synthesized by exploiting various approaches, but the photophysical method utilizing ultrashort laser sources attracted a lot of attention. Virtually any solid material can be transformed into a NP colloid employing laser ablation of the target in liquid [21]. A liquid lowers the heat load on the target, confines the vapour and plasma, and increases the shock pressure on the surface. The ablated material disperses into the liquid and nucleates into the NPs. The fraction of the ablated material is inevitably chemically oxidized or reduced on the surface of the NPs [22]. Cu and its derivative NPs can also be produced by reductive laser ablation of Cu precursor suspensions [23] or by laser fragmentation of micrometre-sized powder in liquid [23, 24].

Under laser ablation of the target in liquid, Cu mainly forms two stable oxides: monoclinic cupric oxide CuO and cubic cuprous oxide $Cu_2O$ [25]. Metastable paramelaconite $Cu_4O_3$ particles were also reported in Cu ablation experiments [26]. M.A. Gondal *et al.* reported two copper phases $Cu/Cu_2O$ in a water solution that were changed into $Cu/CuO$ when a hydrogen peroxide oxidizing agent was present in the liquid medium [27]. A. Nath *et al.* demonstrated that under tightly focused ablation conditions CuO formation with typical 200 nm diameters was observed whereas under defocusing conditions $Cu_2O$ nano colloids of particles smaller than 10 nm were obtained [25].

Cu and some of its oxide NPs are not very stable in a natural environment and are prone to ageing [28]. Therefore, the capping layer of a NP or the use of stabilizing matrix was previously investigated, for example, copper-graphene nanocomposite [3], $Cu_2O$ NP bound with polyurethane [29], $Cu_2O$ nano-colloid using sodium alginate as a capping agent [30]; Cu NPs in polyaniline [31], Cu-NP-containing shellac biopolymer resin [32], Cu NPs in polylactic acid [33], Cu nanowires in Zeolitic imidazolate framework [34, 35].

The reported stability, structure, and crystallinity of laser-ablated Cu NPs are very different depending on the solvent [36-38]. For example, when made in pure water, Cu NPs were least stable, while the ageing of synthesized in ethanol and acetone was slower as seen from the plasmon peak-related colour changes [28]. Many different solvents in addition to water have been investigated, including ethanol [28], acetone [39], decane [26], chloroform [40], superfluid helium [41], etc.

The size distribution of NPs generated by laser ablation in liquid tends to decrease after longer irradiation durations because two competing processes namely growth and fragmentation/melting of the NPs occur [22]. A 50 nm blue shift of the diameter-related localized surface plasmon resonance (LSPR) absorption peak in the Cu NP colloid was observed during the processing duration of 50 s when the metal target was ablated in walnut oil, which was confirmed by the TEM analysis as a decrease in size from 25 to 5 nm [42]. The aspect ratios of the Cu NPs can be controlled by the external DC electrical field resulting in elongated NPs [43].

The reported antiviral efficacy of liquid-suspended Cu and copper iodide [44] NPs appears promising, but in real-life situations copper-based antiviral coatings find more practical applications. A variety of effective Cu film and coating deposition methods indicated well-expressed antiviral activity, including simple drop-casting of Cu colloid on glass substrates [45], Cu composite polymer deposition using a sponge [29], application of Cu-containing paint [46], dip coating and sol-gel synthesis [47], self-assembly of Cu-binding peptide [48], thermal evaporations of the Cu thin film and annealing [35], magnetron sputtering of the Cu thin film after ion beam treatment [49, 50], flame deposition of Cu oxide NPs [51], cold spraying of Cu particles with a high-pressure carrier gas [52, 53], dealloying Mn–40Cu alloy surface [54].

Spray deposition is advantageous in a way that the Cu NP coating can be easily applied both on flat [55] and on elaborated surfaces like textiles [56] also it can be deposited layer by layer [57] forming a functional surface.

In this work, a cost-effective, high throughput method to produce efficient antiviral Cu-NPs-based coatings produced by spray-coating is presented. Femtosecond laser ablation of the Cu target in water with sodium citrate was used for the generation of stable colloidal Cu NPs which were applied as an active material for developing antiviral surfaces. Polyvinyl butyral was utilized as the binding agent for the spray-coated Cu NPs. The virucidal efficacy of coatings containing different Cu NP content was confirmed against animal-derived model viruses, namely RNA-containing coronavirus and deoxyribonucleic acid (DNA) containing herpesvirus.

## 2. Methods

### 2.1. Production and Stabilization of Copper Nanoparticles

Colloidal Cu nanoparticles were produced by laser ablation of a pure copper (99.9%) metal target (acquired from the Lithuanian Mint, Lithuania) in ultrapure type 1 water (18.2 MΩ/cm) with a 0.02 mmol concentration of sodium citrate surfactant (Sigma Aldrich, USA). The metal target was sonicated in acetone to eliminate any organic contamination before the ablation. The target was placed in a glass beaker and filled with 14 ml of an aqueous solution which ensured approximately 4 mm of the liquid layer above the flat surface of the metal target. The laser ablation was carried out employing a laser micromachining workstation (FemtoLab, Altechna R&D, Lithuania) and a 270-fs pulse length, 1030 nm wavelength Yb:KGW femtosecond laser (Pharos, Light Conversion, Lithuania) operating at 200 kHz repetition rate and 3 W average power. The laser beam was scanned with a galvanometer scanner (SCANcube III 14, ScanLab, Germany) and focused down to *ca.* 20 µm spot size by a telecentric 163 mm focal distance f-Theta lens (SillOptics, Germany). The focal plane with respect to the metal target surface was adjusted with a motorized *z* stage (Physik Instrumente, Germany). More details about the used setup are described elsewhere [58]. The sample treatment was controlled with the SCA software (Altechna R&D, Lithuania) and the surface area of 3 mm x 3 mm was scanned by 30 µm separated lines at a 20 mm/s speed repeating the process 300 times. The overall treatment for each batch of colloid lasted *ca.* 79 min. The principal scheme of the laser ablation setup and process is depicted in **Figure 1 a**. The described process was repeated several times exchanging the colloid with water. The resulting overall colloid volume was concentrated, and the solvent was exchanged with isopropanol by employing centrifugation as described in the supplements and depicted in **Figure S1**. The resulting Cu NP colloid was used for spray coating.

## 2.2. Spray Coating

The deposition of the investigated films on the 1 cm x 1 cm soda lime glass microscope slide substrates (Chemland, Poland) was carried out in two steps. After cleaning the substrates in ethanol (99.5%, Emparta, Merk, Germany), they were first spray-coated with Polyvinyl butyral of 3 mg/ml concentration in isopropanol (99.7%, Chempur, Poland). PVB spray coating was carried out with an aerograph AD-776B Airbrush MAR EW-6000B (Adler, Poland) employing a 0.2 mm nozzle (air pressure 2 bar) from a constant 10 cm distance. The effective thickness of the layer was monitored with the custom-made UV light transmission meter described elsewhere [59] and kept at a 3% optical extinction level throughout the samples. The spray-coating process is summarized in **Figure 1 b**. Several PVB-coated glass samples termed "PVB" were used as prepared for antiviral testing as control. Then, in the second step, the Cu NP colloid ink was spray coated on the PVB layer at 1 bar pressure. The concentration of deposited Cu was controlled by varying the duration of deposition and was also monitored with the UV transmission meter indicating optical extinction from 10% to 25%, which correlated with the Cu NP content. The samples were termed "PVB+CuO 10%", "PVB+CuO 15%", and "PVB+CuO 25%".

## 2.3. Physical Characterization of the NPs and the Coating

The optical density of the resulting Cu colloids and their stability over time was measured in the UV-Vis-NIR spectral range employing AvaSpec-2048 (Avantes, the Netherlands) fibre spectrometer of 1.4 nm spectral resolution and combined deuterium and halogen light source AvaLight-DHc (Avantes, the Netherlands).

The Cu NP size distribution was evaluated from field emission gun scanning electron microscope (SEM) Quanta 200 FEG (FEI, Czech Republic) micrographs. Diameters of the nanoparticles were extracted from SEM images using a custom multi-step processing algorithm implemented in MATLAB (MathWorks, USA). Image binarization (separation of particles from their background) was done using adaptive thresholding. Stuck-together particles were segmented using the watershed transformation similar to R. Baiyasi *et al.* [60]. The diameter of each particle was determined from the occupied area assuming that particles are round.

The structure of the NPs was further investigated with the transmission electron microscope (TEM) Tecnai G2 F20 X-TWIN (FEI, Czech Republic). The Cu content in the coatings was obtained with the energy-dispersive X-ray spectrometer system (EDS) Quantax 200 (Bruker, Germany) attached to the SEM under 5 kV accelerating voltage.

The crystal structure of Cu NPs was determined using a D8 Discover X-ray diffractometer (XRD, Bruker AXS GmbH, Germany) operating at 40 kV and 40 mA with Cu K$_\alpha$ ($\lambda$ = 1.5418 Å) radiation source and parallel beam geometry with 60 mm Göbel mirror. Soller slit with an axial divergence of 2.5° was utilized on the primary side. Diffraction patterns were recorded using a fast-counting LynxEye detector with a 2.475° opening angle and 6 mm slit opening. The peak intensities were scanned over the range of 20 – 90° (coupled 2$\theta$-$\theta$ scans) with 0.02° step size.

Raman scattering measurements of Cu NPs were carried out employing an InVia (Renishaw, UK) spectrometer and a 532 nm laser focused with x50 times objective 0.75NA lens. For the analysis, a 50 µl drop of Cu colloidal solution was dried out on a glass substrate.

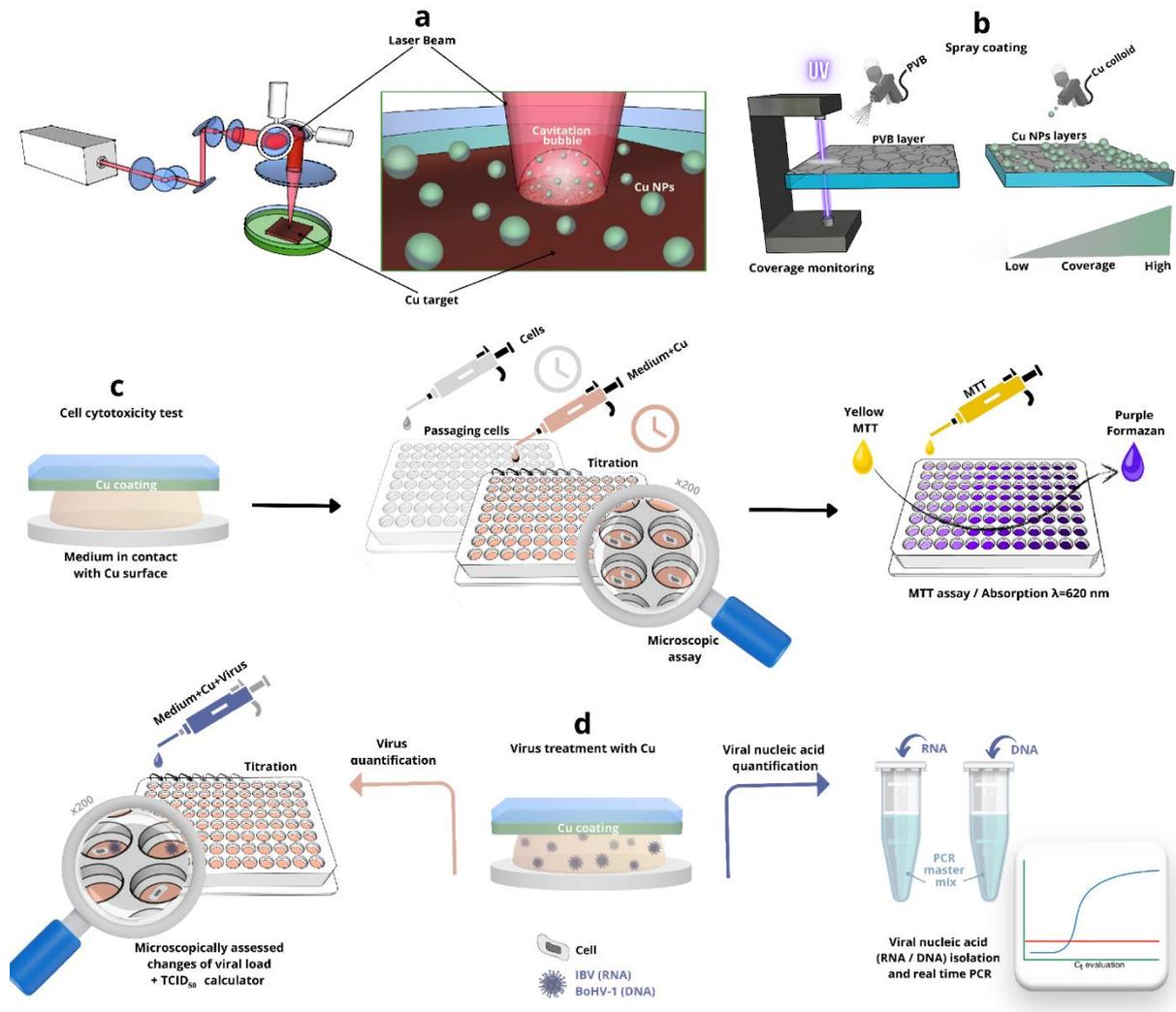

**Figure 1** Preparation of Cu NP loaded coatings and general scheme of the *in-vitro* experiments. (a) Femtosecond laser micromachining setup with a galvanometric scanner and close-up view of the processes in the Petri dish where the Cu target in liquid is scanned with a laser during the ablation process; (b) two-stage spray coating process starting with deposition of PVB which is followed by deposition of Cu-NPs, the UV light transmission meter is used for monitoring effective thicknesses of the coatings; (c) cell cytotoxicity test; (d) the treatment of virus culture by contacting with investigated coatings and preparation of samples for two quantitative analyses, *i.e.*, virus quantification employing calculating tissue culture infectious dose, and viral nucleic acids via real-time polymerase chain reaction.

## 2.4. Antiviral Testing

### 2.4.1. Model Viral Strains and Cell Cultivation

The developed coatings were tested with two types of viruses: (i) animal-origin coronavirus containing RNA, the agent of avian infectious bronchitis (IBV), and (ii) herpesvirus containing DNA, the agent of bovine herpesvirus infection (BoHV-1). Viruses were propagated using two kinds of cell cultures, *i.e.*, Vero and MDBK/NBL-1, respectively.

The Vero-adapted cytopathogenic IBV Beaudette strain was obtained from Dr M. H. Verheije (Utrecht University, The Netherlands). Madin-Darby bovine kidney (MDBK)-adapted, cytopathogenic BoHV-1 4016 strain [61] was provided by Dr I. Jacevičienė (National Food and Veterinary Risk Assessment Institute, Lithuania). The viruses were stored in a deep freezer at -80°C.

Vero (ATCC CCL-81™) and MDBK (NBL-1, Bov.90050801) cell cultures were obtained from Dr I. Jacevičienė (National Food and Veterinary Risk Assessment Institute, Lithuania). Vero and MDBK/NBL-1 cells were cultured in Dulbecco's modified eagle's medium (DMEM, Gibco, UK) and DMEM/Nutrient Mixture F-12 (DMEM / F-12, Gibco, UK), respectively with 10% fetal bovine serum (FBS, Gibco, UK) at 37°C in a 5% $CO_2$ incubator. Nystatin (100 units/ml) and Gentamicin (50 µg/ml) were used to prevent microbial contamination for Vero cells, while Penicillin (100 units/ml) and Streptomycin (100 µg/ml) were used to prevent microbial contamination for MDBK/NBL-1 cell. Depending on the description of the method, the tests were performed at temperatures of 20±2°C (class II biosafety cabinet), and 37°C (cell incubation).

### 2.4.2. Cytotoxicity Control

The cytotoxicity of different Cu content containing coatings was determined in Vero and MDBK cells using MTT assay [62]. First, cells were passaged at a concentration of $1 \times 10^4$ cells/well in a 96-well plate (TPP, Switzerland) and grown at 37°C for 24 h.

Then 10 µl of medium containing 2% FBS was placed on the surface of the 6 well tissue culture test plate (TPP, Switzerland), covered with the surface of the 1 cm × 1 cm sized glass substrates with the investigated coatings, and incubated at room temperature (21±1°C) for 1 h. The thickness of the medium layer was *ca.* 100 µm. After incubation, the liquids were initially diluted 1:50 with a medium containing 2% FBS, and then, employing two-fold dilution steps the volumes of 100 µl were prepared for the treatment of cells. Dilutions from 1:50 to 1:1600 were prepared for the evaluation of cytotoxicity.

Later, the growth medium was removed from cell culture plate wells, cells were washed twice with phosphate-buffered saline (PBS), and treated with 100 µl with titrated liquid at 37°C in a 5% $CO_2$ incubator for 72 h. Before the MTT assay, the cells were microscopically observed and compared with control cells. After 72 h MTT reagent (10 µl, 5 mg/ml, Sigma-Aldrich) was added and incubated for 4 h at 37°C. Then the liquid was carefully discarded and 100 µl dimethyl sulfoxide (DMSO, Carl Roth, Germany) was added to each well and the plates were placed on the shaker for 5 min. Each investigated coating was tested in octuplicate once. The absorbance of each well was measured at 620 nm in a microplate reader (Multiskan™ FC Microplate Photometer, Thermo Fisher Scientific, China) and the percentage of cell survival was calculated. Finally, the dose-response curves were plotted to allow the calculation of 50% cytotoxicity concentrations ($CC_{50}$) that cause lysis and death in 50% of cells [63]. The principle scheme of the investigation is sketched in **Figure 1 c**.

### 2.4.3. Virucidal efficacy testing

### 2.4.3.1 Virus treatment

Suspensions of IBV and BoHV-1 in a volume of 0.01 ml were placed on 6 well tissue culture test plates and covered with investigated different Cu content containing coatings or PVB coated glass plate for control and incubated for 1 hour at room temperature (**Figure 1 d**). After incubation, the liquids were diluted 1:50 with medium containing 2% FBS and were used for the virucidal quantification (the determination of virus titre) in *Section 2.4.3.2* and viral nucleic acid quantification employing real-time PCR as described in *Section 2.4.3.3*.

*2.4.3.2 Virus Quantification*

The virucidal efficacy of the investigated coatings that resulted in a change of viral load was evaluated by determining and comparing the tissue culture infectious dose ($TCID_{50}$) before (virus control) and after exposure on the surfaces of investigated coatings (affected viruses). Cytopathogenic effect (CPE) in sensitive cells was microscopically assessed after 72-120 hours of virus incubation. The principal scheme of the investigation is sketched in **Figure 1 d**. The viral titres of IBV and BoHV-1 were calculated by the Spearman-Karber method using the Marco Binder $TCID_{50}$ calculator [64, 65].

*2.4.3.3 Viral Nucleic Acid Quantification*

*2.4.3.3.1 IBV Real-Time Taqman Reverse Transcription Real-Time Polymerase Chain Reaction*

Real-time TaqMan reverse transcription polymerase chain reaction (RT-PCR) was used to compare the amount of IBV RNA before (control) and after incubation with investigated coatings. The RNA used in the real-time Taqman RT-PCR was extracted employing TRIzol Reagent (Thermo Fisher Scientific, USA) according to the manufacturer's instructions. Real-time RT-PCR was performed as described by Meir [66]. Briefly, a conserved region of 336 base pairs located at nucleotide position 741–1077 of the H120 strain N gene sequence (GenBank accession no. AM260960) was used to design primers and probe for the real-time RT-PCR assay. A forward primer (IBV-f), a reverse primer (IBV-r), and a TaqMan® probe (IBV-p) were used to amplify and detect the 130-bp fragment. Details are provided in **Table S1**. Both the primers and probe were synthesised by Applied Biosystems, UK. PCR amplifications were carried out in a volume of 25.0 μl containing 12.5 μl 2 × RT-PCR buffer mix (AgPath™ OneStep RT-PCR kit, Applied Biosystems), 1 μl 25 × RT-PCR enzyme mix (Applied Biosystems), primers to a final concentration of 400 nM, probe to a final concentration of 120 nM, 2 μl RNA template, and nuclease-free water.

The reaction was carried out in StepOne™ Plus real-time PCR system (Applied Biosystems) at 45°C for 10 min, 95°C for 10 min, and 40 cycles of 95°C for 15 s and 60°C for 45 s. Amplification graphs were recorded and analysed and the threshold cycle (Ct) was determined with Mastercycler RealPlex2 (Eppendorf, Germany) [67]. The principle scheme of the investigation is sketched in **Figure 1 d**.

*2.4.3.3.2 BoHV-1 Real-Time Taqman PCR*
BoHV-1 real-time Taqman PCR was used to compare the amount of BoHV-1 DNA before (control) and after incubation with investigated coatings. DNA isolation was carried out using the Genomic DNA purification kit (Thermo Fisher Scientific, USA, K0721). A protocol was used as described by Lelešius [63].

BoHV-1 forward (BoHV-1-f) and reverse primers (BoHV-1-r) and probe (BoHV-1-p) for quantitative real-time PCR (TaqMan) were designed with Primer Express software (version 1.0; Applied Biosystems, USA) to amplify sequences (product size 97bp) within the open reading frames of the glycoprotein B genes of BoHV-1 [68]. Oligonucleotide primers and MGB-labelled probes were synthesized by Invitrogen (USA) and are detailed in **Table S2**. Amplifications were performed using a TaqMan Universal master mix II (catalogue #4440038, Applied Biosystems, USA). Briefly, real-time PCR amplifications were carried out in a volume of 25.0 μl containing 12.5 μl of master mix, BoHV-1-f and BoHV-1-r (final concentration 240 nM each), BoHV-1-p (final concentration 160 nM), and 10.9 μl of the DNA template. The real-time PCR conditions for the reactions were set as follows: 2 min at 50°C, 10 min at 95°C, and then 40 cycles consisting of a denaturation step at 95°C for 15 s and an annealing-elongation step at 60°C for 1 min.

Amplification plots were recorded and analyzed, and Ct was determined using a Mastercycler (Eppendorf). The real-time PCR was repeated four times and Ct values were recorded.

### 2.5. Cu release into the medium

An identical set of samples were immersed in a 2 ml DMEM cell maintenance medium for 1 hour at room temperature. The concentrations of the Cu released from the coatings into the medium were determined using a double-beam atomic absorption spectrometer AAnalyst 400 (Perkin Elmer, USA). Quantitative analysis of the Cu amount was carried out using the calibration curve that was determined by measuring reference solutions. Afterwards, the optical density of the test solutions was measured and the concentration of the test element in the solution was determined from the calibration curve.

### 2.6. Statistical analysis

The results were analysed with Origin Lab 2023 and the mean comparative studies of NP sizes, $TCID_{50}$, and Ct estimation were done by two-way analysis of variance (ANOVA) with the quantitative data expressed as the mean ± standard deviation (SD). A statistical significance was considered at *$p$ <0.05. Linear regression and the square of the Pearson product-moment correlation coefficient ($R^2$) were used for the evaluation of the correlation between Cu content and cytotoxicity, virus titre, and PCR Ct values.

## 3. Results

### 3.1 Cu NPs.

After the femtosecond laser ablation of the Cu target in liquid, the translucent colour of the solvent changed into pale green (**Figure 2 a**, inset) indicating the formation of the Cu NP colloid. The absorption peak visible in the optical density spectra at *ca.* 620 nm is related to the LSPR of Cu NPs (**Figure a**). The bigger amplitude peak at *ca.* 300 nm is addressed to the oxidized Cu [38,69,70].

Without the addition of surfactant, Cu colloid NPs tended to oxidize and change colour into pale yellow days after the process (results not shown here). Andal and Buvaneswari [30] reported similar results in the colour change where the presence of $Cu_2O$ was confirmed by the XRD method. In this work, the XRD diffractogram indicated similar peaks but of different intensities supporting the domination of metal Cu in a mixture with $Cu_2O$ (**Figure 2 f**). On the contrary, some others claimed to have green appearing laser synthesized CuO [71] but Cu and Cu oxides are not stable and tend to change within hours as reported for chemically synthesized CuO that are prone to transform in $Cu_2O$ [72]. Therefore, Raman scattering measurements were carried out to confirm the material composition as the Raman activity is related to the function of the space group symmetry of a crystalline solid [38]. The spectra obtained with peaks characteristic for CuO and $Cu_2O$ as described in [73-75] are shown in **Figure 2 c** and detailed in **Table S3**. The presence of oxygen in Cu NPs was also confirmed by the EDS mapping of the drop-casted Cu NP colloid on the crystalline silicon surface (**Figure S2**).

The size distribution of the Cu NPs obtained from the SEM micrograph (**Figure 2 e**) is depicted in **Figure 2 d**. The mean diameter of the NPs was 32±14 nm. The shape of the Cu NPs can be assessed better in the TEM micrograph (**Figure 2 e**) and it resembles spheres, which is expected for laser-ablated NPs.

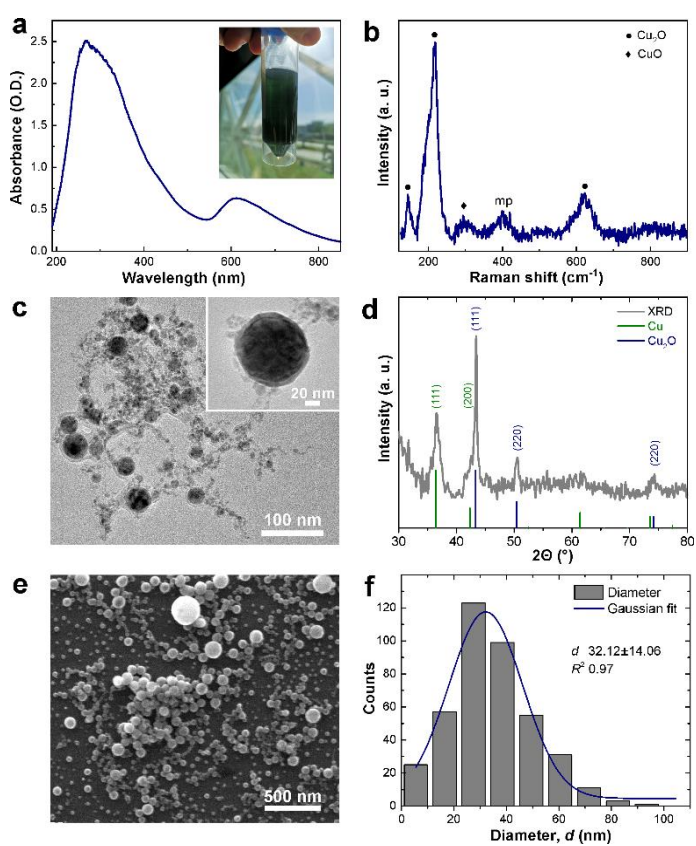

**Figure 2** Analysis of the Cu colloid. (a) UV-Vis-NIR absorbance (optical density, O.D.) spectra of the concentrated colloid. The inset shows a camera image of the concentrated Cu colloid. (b) Raman scattering spectra of the CuNPs with the identified phases addressed in **Table S3**, "mp" – multiphoton, "a.u." – arbitrary units. (c) Micrograph of the Cu NPs drop-casted on a silicon substrate. (d) Size distribution of Cu NPs obtained from the SEM micrographs. (e) TEM micrograph of the Cu NPs. (f) XRD of the Cu NPs with the indicated *hkl* indexes for Cu and $Cu_2O$.

*3.2 Coatings Containing Copper NPs*

The Cu NP ink was spray-coated on PVB-coated glass substrates to monitor the UV transmittance and therefore qualitatively control the effective thickness of the deposited coating. The Cu content in the coatings was evaluated quantitatively by energy-dispersive X-ray spectroscopy (EDS). The EDS spectra are depicted in **Figure S3** where one can see an emergence of the Cu-related $L_\alpha$ peak with increasing spray coating duration. This correlates with the increase in UV extinction. A complete list of detected elements and their concentrations including the substrate is provided in Supplementary **Table S4**. The normalized elemental composition of the coating, obtained after subtracting the elements attributed to the glass substrate and leaving only Cu which is in the form of NPs, and C which is mainly related to PVB coating, is presented in **Table 1**.

**Table 1** Normalized elemental coating composition in weight percent (wt.%) and atomic percent (at.%) together with released Cu concentrations in cell medium after 1-hour immersion. The full coating composition is provided in **Table S4**

| Sample No. | | Element concentration | | Released Cu mg/L |
|---|---|---|---|---|
| | | C | Cu | |
| PVB+CuO 10% | at.% | 97.1 | 2.9 | 0.377 |
| | wt.% | 86.3 | 13.7 | |
| PVB+CuO 15% | at.% | 92.8 | 7.2 | 0.673 |
| | wt.% | 71.0 | 29.0 | |
| PVB+CuO 25% | at.% | 88.8 | 11.2 | 1.972 |
| | wt.% | 59.9 | 40.1 | |

Differences in surface morphology are summarized in **Figure 3.** PVB appears as a random mesh that stems from the droplets due to the chosen spray coating method as seen on the micrometre range scale magnification (**Figure 3**, first column). At the nanometre scale level magnification, Cu NPs are seen either embedded or found in the valleys between PVB mesh (**Figure 3,** third column).

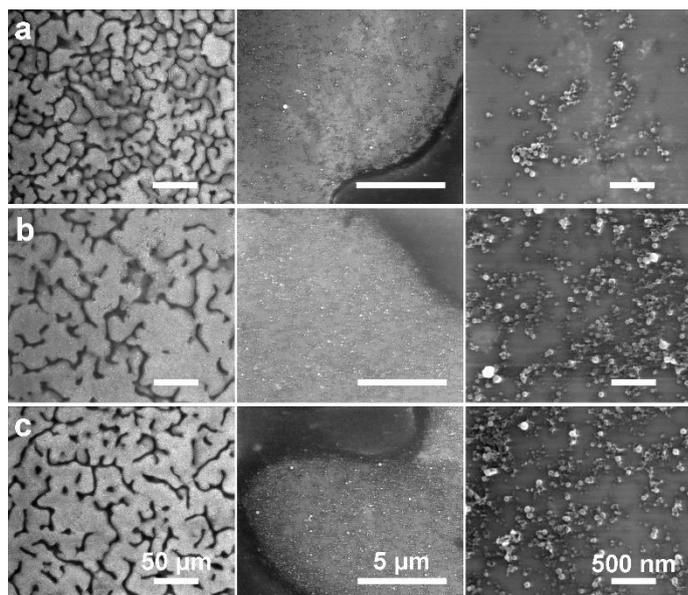

**Figure 3** SEM micrographs of PVB coatings with different Cu NP content, (a) PVB+CuO 10%, (b) PVB+CuO 15%, (c) PVB+CuO 25% at characteristic magnifications of 1k x, 20k x,

100k x summarized in first, second, and third columns, respectively, with the scale bars of identified length.

## 3.3. Cytotoxicity Control

Cytotoxicity control was performed to differentiate cytotoxic and noncytotoxic Cu concentrations for Vero and MDBK cells and choose the correct concentrations for further virus treatment. It was obtained that PVB+CuO 10%, PVB+CuO 15%, and PVB+CuO 25% coatings were cytotoxic for both cell lines and caused the death of 50% of cells under dilutions identified in **Table 2**. A similar experiment where identical samples were immersed in a 2 ml medium resulted in particle per million (ppm) level Cu concentrations (**Table 1**), which recalculated into the µg/µl concentrations translates into 377.0 µg/µl to 1972.0 µg/µl. It was found that the cytotoxicity correlated linearly with the content of Cu in the coatings while linear correlation with the Cu released in the medium was less expressed. Virucidal assays were carried out with higher dilutions than identified in **Table 2** ensuring that cytotoxicity from investigated films is not prevailing and is not influencing the virucidal assay.

**Table. 2** Identified cytotoxic dilutions of investigated copper nanoparticle-based coatings for Vero and MDBK cells. $R^2$ represents the correlation between the Cu concentration in the film at.% or released in the medium and determined cytotoxic dilutions.

| Cells | Cytotoxicity, $CC_{50}$ (dilution) | | | $R^2$ | |
|---|---|---|---|---|---|
| | PVB+CuO 10% | PVB+CuO 15% | PVB+CuO 25% | Cu in film | Cu released |
| Vero | 1:32.0 | 1:59.0 | 1:74.6 | 0.90 | 0.60 |
| MDBK | 1:25.8 | 1:49.7 | 1:69.5 | 0.93 | 0.64 |

## 3.4. Virucidal efficacy

The change in the number of two investigated type viruses after 1 h of contact incubations with different investigated coatings expressed as a decrease in $\log_{10}$ $TCID_{50}$/ml, is depicted in **Figure 4 a** and **Table S5**. The Cu NP content in the coating had a direct influence on the virus activity. All concentrations had a significant reduction effect on IBV and BoHV-1 titre from 2.24 to 5.00 $\log_{10}$ $TCID_{50}$/ml and 1.87 to 3.38 $\log_{10}$ $TCID_{50}$/ml, respectively (**Figure 4 a**). Samples of the highest Cu NP content coating (PVB+CuO 25%, or 11.2 at.% of Cu) inactivated all coronaviruses (100%), while for herpesviruses 99.96% reduction was obtained. Linear regression of the Cu content in the film with the $TCID_{50}$ $\log_{10}$ virus reduction results in a coefficient of determination $R^2 = 0.80$ and $= 0.99$, for IBV and BoHV-1, respectively while calculating with the Cu content released in the medium $R^2 = 0.99$ and $= 0.94$.

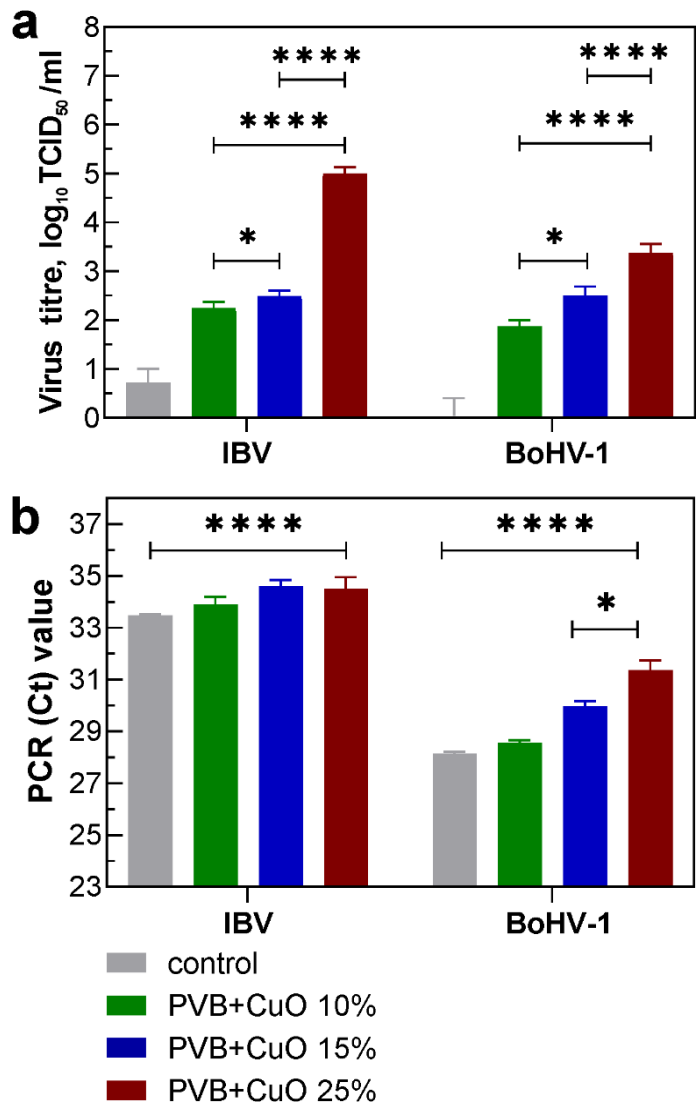

**Figure 4.** *In vitro* study of different PVB+CuO (10%, 15%, and 25%) coatings and control, (a) virus titre reduction $\log_{10}$ of IBV, and BoHV-1 after 1 h of contact (*$p$ = 0.0117 and ****$p$ = <0.0001) and PVB control. (b) Cycle threshold (Ct) values the virus control and coatings of the RT-PCR (*$p$ = 0.0042 and ****$p$ = <0.0001).

*3.5. Real-Time PCR*

The change in the number of nucleic acids of IBV and BoHV-1 after contact with the coating samples was evaluated with real-time PCR by determining the threshold cycle Ct+ limit value increase from the PCR cycle amplification plots. The cycle threshold values depicted in **Figure 4 b** indicate that for IBV (**Table S5**) the PCR Ct values varied minimally from 33.5 to 34.6. While for BoHV-1 significant Ct increase was registered from 28.6 to 31.4. Linear regression of the Cu content in the film with the Ct threshold value results in a coefficient of determination $R^2$ = 0.65 and 1.00, for IBV and BoHV-1, respectively while calculated with the Cu released in the medium $R^2$ = 0.29 and 0.88.

**4. Discussion**

The presence of the $Cu_2O$ phase in the Cu NPs confirmed by XRD, EDS, and Raman studies indicated that the laser-ablated Cu colloids tend to oxidize, but they did not change their characteristic green colour related to the LSPR of the Cu NPs in time. The presence of sodium citrate in water helped to prevent the fast ageing of the Cu colloid that was observed by others [38]. Exponentially increasing Cu release in the cell medium on the Cu content in the coating suggests that higher Cu NP loads in the coating have poorer binding with the surface. It might be related to longer deposition duration and therefore dissolving of the binding PVB agent. Similar nanocomposite coatings were investigated by Toledo *et. al* [76] where commercial Cu and CuO nanoparticles were introduced into poly(methyl methacrylate) and polyepoxide matrixes and therefore resulted in an antiviral effect on human coronavirus HCoV-OC43.

The two-stage spray coating process enabled effective deposition and control of the Cu content on PVB-coated glass substrates and could be a more controlled alternative for dip coating and drop-casting which were also used in similar studies [45, 47]. It was not investigated in more detail here but the used setup can coat complex surfaces, for example, various touch surfaces [53] and textiles [56] because it is not limited by direct visibility like in fixed-source vacuum deposition systems but rather can access the surface of any pre-existing part utilizing a robotic arm or a handheld applicator.

Cu NPs-containing coating antiviral studies were started from the cytotoxicity testing aiming to assess their potentially destructive effects on different cell systems before evaluation of their virucidal effects. The testing on Vero and MDBK cell cultures revealed a similar impact on the viability of cells. Our study showed that the cytotoxicity has a direct linear correlation with the Cu content in the coatings and is also related to the Cu released in the medium. The Cu content release rate in 2 ml medium volume might be different compared to the original experiment where a 10 µl drop was spaced in a 100 µm thickness layer and therefore the initial Cu concentration gradients might have saturated faster. Similar data obtained from the MTT assay indicated a strong dose-response relationship with respect to copper toxicity [77]. Reduced cell viability at longer contact durations or bigger Cu concentrations can be related to the loss of membrane barrier or cell surface microvilli that could be caused by oxidation stress [78].

Cu NP coatings demonstrated expressed virucidal efficacy on both investigated RNA and DNA viruses. The reduction in the number of viruses after 1-hour contact correlated with the atomic Cu concentration in the coatings and it was more pronounced for the IBV than for the BoHV-1. While correlation with released Cu was more pronounced for IBV. That suggests that some differences in the Cu NP virucidal effect exist for investigated RNA and DNA viruses. For the inactivation of IBV contact with the coating might be the prevailing mechanism while for BoHV-1 both, the contact with the bonded NPs and the released Cu ion and Cu NPs might play the role. More detailed studies are necessary to support this hypothesis and understanding of different mechanism impact and evolution in time. The absolute highest decrease in IBV biological activity of 100.0% was observed after contact with PVB+CuO 25% sample throughout the entire scope of the study.

The real-time PCR results indicated different quantitative effects on investigated viruses and their nucleic acids. The Ct threshold values were following the concentration trends only for the BoHV-1 virus indicating the Cu NP effect on DNA damage. It is important to note, that PCR detects intact specific target sequences of viral nucleic acids of both alive and inactivated viruses. Therefore, assessing the comparative percentage composition of nucleic acids of live and inactivated viruses in a sample by PCR alone is not possible. Previous studies showed that Cu NPs can have a direct degenerative effect on the biological structures of viruses, viral RNA, and short RNA fragments can be detected in post-contact titration and culture media of viruses [46, 79].

Our IBV real-time PCR confirmed the results of previous research and showed that Ct values are not the markers for infectivity of coronavirus [80]. It was proved by both studies when RNA sequences were detected after inactivation of 100% coronaviruses. The comparison of the results of the quantification of viruses and their nucleic acids showed a higher decrease in virus infectivity than in the degradation of nucleic acids. The data from our study with BoHV-1 also showed that the infectious titre of the viruses (**Figures 4 a**) after 1 hour of contact with the surfaces decreased significantly faster than the DNA damage detected by PCR (**Figures 4 b**). This could be explained by the fact that nucleic acids are significantly more resistant to external influences than viral proteins or envelope lipids [81].

In our study, the high antiviral efficacy indicated by the virus titres can be addressed by the fact that the viruses were not resistant to Cu which caused irreversible virus activity and possible RNA and DNA morphological and structural alterations. Multiple mechanisms having an internal effect on the virus were identified in literature including (i) the production of reactive oxygen species (ROS) through free copper ions ($Cu^+$) released from the NP, leading to the denaturation of deoxyribonucleic acid (DNA)/ ribonucleic acid (RNA) and damaged virion integrity [13, 51, 82-84]; (ii) $Cu^+$-induced virus inactivation by oxidizing lipids can inactivate viruses leading to the degradation of virus proteins through the generation of hydroxyl radicals [13, 85]; (iii) cytotoxicity caused by free radicals that interact with the components of the virus protein and generating hydroxyl radicals of transition metal hydroxyl radicals bound to the proteins [13, 86]; (iv) the inactivation of viral metalloproteins by replacing the respective metal with Cu [87]; (v) disruption of the capsid integrity of virus and destruction of DNA or RNA genomes by Cu binding and cross-linking between and within strands [84, 88-91]. The external antiviral properties of Cu NPs are related to (i) the "contact killing" [91] and inactivation of the virus through membrane depolarization/ or by some not yet explained mechanisms related to the dissolution of metal ions ($Ag^+$, $Cu^{2+}$, and $Zn^{2+}$) [51, 84]; (ii) surface-related catalytic activity of copper oxides (direct interaction with the surface) [84, 88, 91], (iii) the interference with the virus capability to attach and enter target cells [92, 93] by rapidly damaging the virus surface proteins and membrane, breaking the envelope or losing the virus capacity of self-containing folding upon itself [83, 94].

Since monovalent copper ions ($Cu^+$) are more toxic than divalent ($Cu^{2+}$), these ions might act as a catalytic cofactor for the formation of intracellular ROS. As a result of such exposure, the defragmentation of viral RNA occurs, and short RNA fragments can be detected by PCR [46, 79].

The external S proteins are very important for the protection of the genomic materials of coronaviruses in the external environment and are necessary for the first two stages of the viral biological cycle – "attachment" and "entry" into the host cell [95]. Due to the sensitivity of external morphological structures, the inactivation of enveloped viruses is even faster compared to non-enveloped viruses [13, 96].

Cu concentration is a key determinant of antimicrobial performance, with surfaces containing 55 to 70% active Cu effectively eliminating pathogenic microorganisms, including human immunodeficiency virus, within a very short contact time [97].

Also, it has been demonstrated, that solid copper oxide has been proven to effectively inactivate the influenza virus during contact with $Cu_2O$, inhibiting the functional capacity of the HA protein (one of the main surface antigens) and destroying its biochemical structures [45]. So, our research confirmed that the antiviral effect of Cu-based NPs is universal in different virus systems. In addition, the size of Cu NPs is another important criterion for virucidal efficiency, where smaller nanoparticles were reported to have better activity [98, 99].

## 5. Conclusions

It was demonstrated that copper target ablation in water with 0.02 mmol sodium citrate employing a femtosecond laser can stabilize the Cu nanoparticles and protect them from rapid ageing. Optical extinction control of the concentrated Cu NP ink spray coating on PVB-coated glass allowed varying the copper content from 2.9 at.% to 11.2 at.%. Raman, EDS, and XRD studies confirmed that 32 nm mean-size nanoparticles are mixtures of mainly metallic copper and copper (I) oxide, $Cu_2O$.

Studies of the virucidal activity of RNA-containing coronavirus IBV and DNA-containing herpesvirus BoHV-1 in cell cultures after 1-hour contact with the investigated coatings indicated a definite Cu content dependant negative effect on the biological activity of both model viruses leading to virus inactivation and viral nucleic acid degradation.

The absolute highest decrease in IBV biological activity of 100.0% was observed after contact with 11.2 at.% Cu content coating releasing up to 1.972 mg/L Cu and its compounds in the medium throughout the entire scope of the study.

The real-time PCR results indicated different quantitative effects on investigated IBV and BoHV-1 and their nucleic acids. The Ct threshold values were following the concentration trend for BoHV-1 indicating the Cu NP effect on DNA damage.

## Author Contributions

Conceptualization and methodology – T.T., A.Š., S.R.; Investigation – A.T., R.M., M.A.B., K.D., B.A., R.L., A.Š; Data analysis R.L., A.Š., D.Z., S.B., M.A.B., T.K., A.T., T.T.; Visualization: S.B., R.M., T.T.; Writing – original draft D.Z., S.B., T.T.; Writing – review & editing – D.Z., S.B., R.L., A.Š., S.T., T.T.; Supervision – T.T. and A.Š.; Project administration, Funding acquisition – T.T.

All authors have read and agreed to the published version of the manuscript.

## Funding

This project has received funding from European Regional Development Fund (project No. 13.1.1-LMT-K-718-05-0018) under a grant agreement with the Research Council of Lithuania (LMTLT). Funded as the European Union's measure in response to the COVID-19 pandemic.

## Acknowledgements

A special thanks go to project No. 13.1.1-LMT-K-718-05-0018 members G. Khatmy, M. Ilickas, Dr R. Zykus, Dr A. Urbas, M. Mikutis, O. Ulčinas, J. Baltrukonis, E. Nacius, M. Vainoris, D. Mazur from the Kaunas University of Technology, Lithuanian University of Health Sciences, and Altechna R&D for their technical assistance.

**Conflicts of Interest**

The authors declare no conflict of interest.

**Data Availability**

The datasets used or analysed during the current study are available from the corresponding author upon reasonable request.

# Supplementary Information

## Virucidal Efficacy of Laser-Generated Copper Nanoparticle Coatings Against Model Coronavirus and Herpesvirus

Shahd Bakhet[1], Rasa Mardosaitė[1], Mohamed Ahmed Baba[1], Asta Tamulevičienė[1,2], Brigita Abakevičienė[1,2], Tomas Klinavičius[1], Kristupas Dagilis[2], Simas Račkauskas[1], Sigitas Tamulevičius[1,2], Raimundas Lelešius[3,4], Dainius Zienius[3,4], Algirdas Šalomskas[4], Tomas Tamulevičius[1,2]*

[1]Institute of Materials Science of Kaunas University of Technology, K. Baršausko St. 59, LT-51423, Kaunas, Lithuania

[2]Department of Physics, Kaunas University of Technology, Studentų St. 50, LT-51368, Kaunas, Lithuania

[3]Department of Veterinary Pathobiology, Lithuanian University of Health Sciences, Tilžės St. 18, LT-47181 Kaunas, Lithuania

[4]Institute of Microbiology and Virology, Lithuanian University of Health Sciences, Tilžės St. 18, LT-47181 Kaunas, Lithuania

*Corresponding author: T. Tamulevičius, tomas.tamulevicius@ktu.lt, Tel: +370 (37) 313432, Institute of Materials Science of Kaunas University of Technology, K. Baršausko St. 59, LT-51423, Kaunas, Lithuania

In total 4 tubes with 15 ml of initial colloidal solution were centrifuged for 20 min at 6000 x g (Lace 16, COLO, Slovenia), then the water was removed, and sediments were redispersed in 10 ml of isopropanol (**Figure S1**). The second centrifugation proceeded in the same way, only for the redispersion of sediment, only 5 ml of isopropanol was used to concentrate the final solution that was used for spray coating.

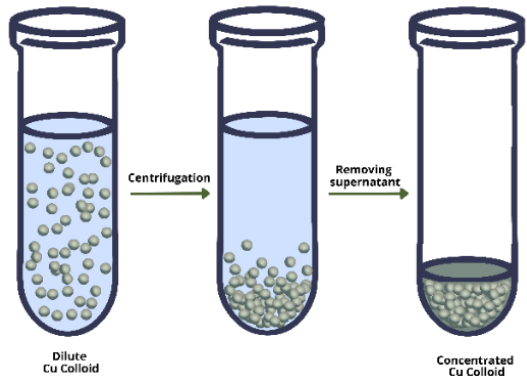

**Figure S1** Concentration of the 15 ml of Cu colloid water via centrifuging and exchange of solvent from water to isopropanol. The last two steps were repeated twice using 10 ml and 5 ml of isopropanol.

The forward primer (IBV-f), a reverse primer (IBV-r), and a TaqMan® probe (IBV-p) used in real-time Taqman reverse transcription polymerase chain reaction (PCR) are detailed in **Table S1**.

**Table S1** The primers and probe for IBV real-time Taqman reverse transcription PCR

| Oligonucleotide | Sequence (5′-3′) | Nucleotide position |
|---|---|---|
| IBV-f | ATGCTCAACCTTGTCCCTAGCA | 811–832 |
| IBV-r | TCAA-ACTGCGGATCATCACGT | 921–941 |
| IBV-p | FAM-TTGGAAGTAGAGTGACGCCCA AACTTCA-BHQ1 | 848–875 |

The forward (BoHV-1-f) and reverse primers (BoHV-1-r) as well as the probe (BoHV-1-p) for real-time TaqMan PCR are detailed in **Table S2**.

**Table S2** The primers and probe for BoHV-1 real-time TaqMan PCR

| Oligonucleotide | Sequence (5′-3′) | Nucleotide position |
|---|---|---|
| BoHV-1-f | TGTGGACCTAAACCTCACGGT | 57499–57519 |
| BoHV-1-r | GTAGTCGAGCAGACCCGTGTC | 57595–57575 |
| BoHV-1-p | AGGACCGCGAG TTCTTG CCGC | 57525–57545 |

The Raman peaks of Cu NPs related to the function of the space group symmetry of Cu oxides are provided in **Table S3**.

**Table S3** Cu NP Raman scattering peak positions and their assignment for space groups

| No. | Peak position (cm$^{-1}$) | Assignment |
|---|---|---|
| 1 | 150 | $F_{1U}$ ($Cu_2O$) [1] |
| 2 | 216 | 2 $E_U$ ($Cu_2O$) [2] |
| 3 | 296 | $A_g$ (CuO) [2] |
| 4 | 405 | Multiphoton process [1, 3] |
| 5 | 620 | $F_{1U}$ ($Cu_2O$) [1] |

The full elemental composition of the PVB and PVB+CuO (10%, 15%, 25%) coatings of glass substrates obtained with EDS are provided in **Table S4**.

**Table S4** Elemental composition of the PVB and PVB+CuO coatings on the glass substrates in atomic percent (at.%).

|  | Element | PVB | PVB+CuO 10% | PVB+CuO 15% | PVB+CuO 25% |
|---|---|---|---|---|---|
| **Coating** | Carbon (C) | 45.33 | 28.62 | 21.25 | 20.34 |
|  | Copper (Cu) | - | 0.86 | 1.64 | 2.57 |
| **Substrate** | Oxygen (O) | 42.22 | 52.89 | 57.20 | 57.28 |
|  | Sodium (Na) | 1.50 | 2.15 | 2.53 | 2.50 |
|  | Magnesium (Mg) | 0.39 | 0.47 | 0.52 | 0.62 |
|  | Aluminium (Al) | 0.16 | 0.24 | 0.17 | 0.21 |
|  | Silicon (Si) | 8.36 | 12.33 | 14.18 | 14.15 |
|  | Calcium (Ca) | 2.04 | 2.44 | 2.44 | 2.33 |
|  |  | 100.00 | 100.00 | 100.00 | 100.00 |

Elemental maps of drop-casted Cu NPs on silicon substrate are depicted in **Figure S2**.

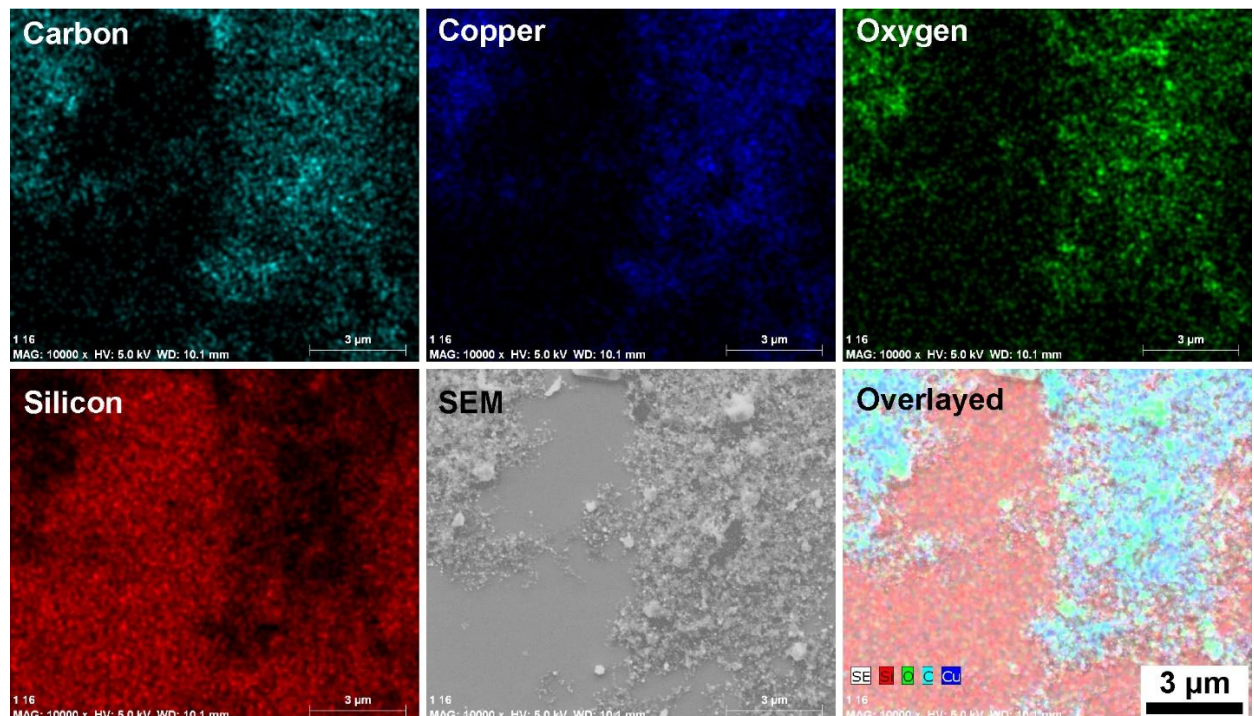

**Figure S2** EDS mapping analysis of the Cu NPs drop-casted on a crystalline silicon surface. The detected chemical elements, namely carbon, copper, oxygen, and silicon are indicated on the maps along with the SEM micrograph and overlaid elemental distribution. The measurements were obtained with a 5 kV accelerating voltage.

EDS spectra of the PVB and PVB+CuO (10%, 15%, 25%) coatings are depicted in **Figure S3**.

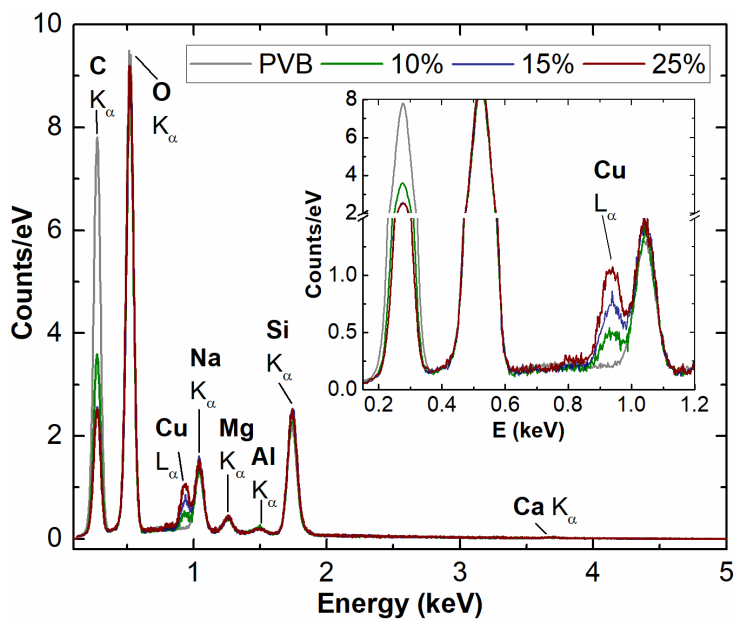

**Figure S3** EDS spectra of PVB spray-coated glass and PVB loaded with different Cu NP content PVB+CuO 10%, PVB+CuO 15%, PVB+CuO 25% coatings. The measurements were obtained with a 5 kV accelerating voltage. The inset highlights the characteristic of the energy range for carbon and copper that are attributed to the coating. The characteristic X-ray peaks are noted with chemical elements along with the Siegbahn transition notation.

Detailed results of IBV and BoHV-1 titres and real-time PCR are provided in **Table S5** and **Table S6**, respectively.

**Table S5** IBV strain "Beaudette" virus biological activity evaluation after 1-hour contact with different Cu content coatings. 10 μl of the virus solution was used, and contact duration with the investigated surface was 1 hour, temperature 20±2°C. Washed for 1 min with 490 μl DMEM (initial dilution – 1:50).

| Sample | Viral titre, $\log_{10}$ $TCID_{50}$/ml, M±σ | Residual viruses, $TCID_{50}$ units/ml | A decrease in the number of viruses | | PCR $C_t$ value, M±σ |
|---|---|---|---|---|---|
| | | | $\log_{10}$ $TCID_{50}$ /ml | % | |
| IBV Control | 5.00±0.13 | 100 000 | 0 | 0 | 33.5±0.02 |
| PVB Control | 4.40±0.29 | 25 100 | 0.60 | 74.89 | 33.0±0.50 |
| PVB+CuO 10% | 2.76±0.13 | 575 | 2.24 | 99.42 | 33.9±0.31 |
| PVB+CuO 15% | 2.51±0.12 | 323 | 2.49 | 99.68 | 34.6±0.24 |
| PVB+CuO 25% | - | - | 5.00 | 100.00 | 34.5±0.46 |

**Table S6** BoHV-1 strain "4016" virus biological activity evaluation with different Cu content coatings. 10 µl of the virus solution was used, and contact duration with the investigated surface was 1 hour, temperature 20±2°C. Washed for 1 min with 490 µl DMEM/F12 (initial dilution – 1:50).

| Sample | Viral titre, $\log_{10}TCID_{50}$/ml, M±σ | Residual viruses, $TCID_{50}$ units/ml | A decrease in the number of viruses | | PCR $C_t$ value, M±σ |
| --- | --- | --- | --- | --- | --- |
| | | | $\log_{10} TCID_{50}$/ml | % | |
| BoHV-1 Control | 7.20±0.0 | 15 848 931 | 0 | 0 | 28.17±0.05 |
| PVB Control | 7.15±0.23 | 14 125 375 | 0.05 | 10.87 | 27.80±0.33 |
| PVB+CuO 10% | 5.33±0.13 | 213 796 | 1.87 | 98.65 | 28.56±0.11 |
| PVB+CuO 15% | 4.70±0.19 | 50 118 | 2.50 | 99.70 | 29.97±0.21 |
| PVB+CuO 25% | 3.82±0.18 | 6 607 | 3.38 | 99.96 | 31.36±0.38 |

**Highlights**

- Laser-ablated Cu/Cu$_2$O nanoparticles in water with sodium citrate are not ageing.
- Cu compounds in spray-coated films have a virucidal effect on model RNA and DNA viruses.
- A 100% decrease of model RNA virus was obtained for 11.2 at.% Cu content coating.
- Real-time PCR results indicated different Cu compound's effects on investigated viruses.

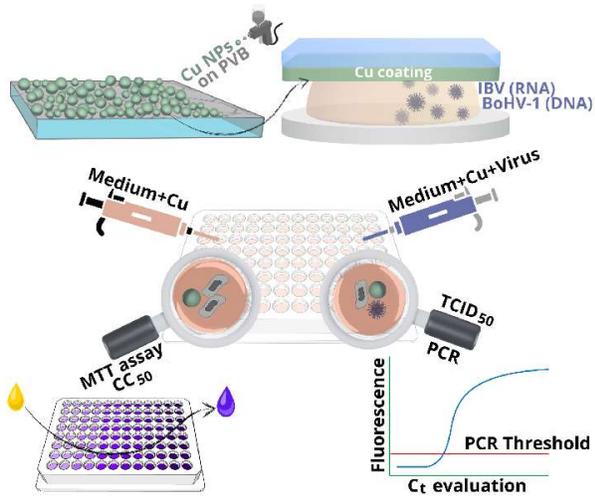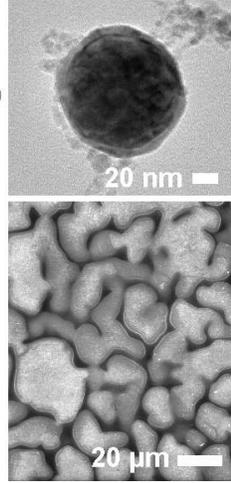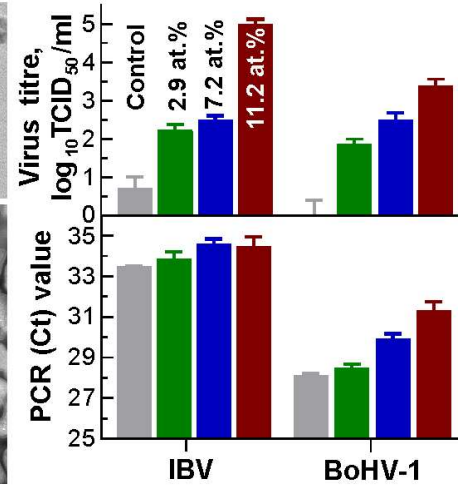